\documentclass[aps,prb,twocolumn,showpacs]{revtex4}
\usepackage{amssymb}

\usepackage{graphicx}

\begin{document}

\title{Observation of a van Hove singularity and implication for strong coupling induced Cooper pairing in KFe$_{2}$As$_{2}$}

\author{Delong Fang$^{1}$, Xun Shi$^{2}$, Zengyi Du$^{1}$, Pierre Richard$^{2,3}$, Huan Yang$^{1}$, X.X.Wu$^{2}$, Peng Zhang$^{2}$, Tian Qian$^{2}$, Xiaxin Ding$^{1}$, Zhenyu Wang$^{2}$, T. K. Kim$^{4}$, M. Hoesch$^{4}$, Aifeng Wang$^{5}$, Xianhui Chen$^{5,6}$, Jiangping Hu$^{2,7}$, Hong Ding$^{2,3,\dag}$, Hai-Hu Wen$^{1,6,*}$}

\affiliation{$^1$National Laboratory of Solid State Microstructures and Department of Physics, Nanjing University, Nanjing 210093, China}
\affiliation{$^2$Institute of Physics and National Laboratory for Condensed Matter Physics, Chinese Academy of Sciences, Beijing 100190, China}
\affiliation{$^3$Collaborative Innovation Center of Quantum Matter, Beijing, China}
\affiliation{$^4$Diamond Light Source, Harwell Campus, Didcot, OX11 0DE, United Kingdom}
\affiliation{$^5$Hefei National Laboratory for Physical Sciences at Microscale and Department of Physics, University of Science and Technology of China, Hefei 230026, China}
\affiliation{$^6$Collaborative Innovation Center of Advanced Microstructures, Nanjing University, China}
\affiliation{$^7$Department of Physics, Purdue University, West Lafayette, Indiana 47907, USA}

\begin{abstract}
Scanning tunneling spectroscopy (STS) and angle-resolved photoemission spectroscopy (ARPES) have been investigated on single crystal samples of KFe$_{2}$As$_{2}$. A van Hove singularity (vHs) has been directly observed just a few meV below the Fermi level (\emph{E}$\mathrm{_F}$) of superconducting KFe$_{2}$As$_{2}$, which locates in the middle of the principle axes of the first Brillouin zone. The majority of the density-of-states at \emph{E}$\mathrm{_F}$, mainly contributed by the proximity effect of the saddle point to \emph{E}$\mathrm{_F}$, is non-gapped in the superconducting state. Our observation of nodal behavior of the momentum area close to the vHs points, while providing consistent explanations to many exotic behaviours previously observed in this material, suggests Cooper pairing induced by a strong coupling mechanism.
\end{abstract}


\maketitle
\section{Introduction}

A two-dimensional saddle point in the electronic band dispersion of a material causes a divergence in the density-of-states (DOS) called van Hove singularity (vHs). Electronic properties are usually expected to vary drastically as a vHs approaches the Fermi level (\emph{E}$\mathrm{_F}$). Enhancement of superconductivity due to a low-energy vHs is well understood for conventional BCS superconductors since the transition temperature (\emph{T}$\mathrm{_c}$) is positively correlated to the electronic DOS at \emph{E}$\mathrm{_F}$ in the case of weakly attractive electron-phonon interactions. However, the impact of a vHs in the unconventional superconductors is much more complex and not well understood.
Theoretically, the presence of a vHs\cite{vHs} near \emph{E}$\mathrm{_F}$ can play an important role in controlling the interplay between superconductivity, spin-density-wave and charge-density-wave, and in inducing exotic chiral superconductivity such as time-reversal symmetry breaking pairing. Such a theoretical approach has been applied to cuprates\cite{cuprates}, doped graphene\cite{graphene} and BiS$_2$-based superconductors\cite{BiS2_1,BiS2_2}. Experimentally, an extended saddle point has been observed in cuprates by angle-resolved photoemission spectroscopy\cite{cuprates_ARPES_saddle}. However, a divergent vHs in DOS, measured by scanning tunneling spectroscopy, is absent in many cuprates, except on some surface areas of highly overdoped Bi$_2$Sr$_2$CuO$_{6+\delta}$ (Bi2201), where the superconducting (SC) pairing vanishes\cite{Bi2201_STM}. A STS study on twisted graphene layers found a vHs in the vicinity of \emph{E}$\mathrm{_F}$\cite{graphene_STM}, prompting the theoretical prediction of chiral superconductivity in this system\cite{graphene}.

KFe$_2$As$_2$ is a stoichiometric end-member of the hole-doped Ba$_{1-x}$K$_x$Fe$_2$As$_2$ family. While its superconducting transition temperature is low (\emph{T}$\mathrm{_c}$ $\approx$ 3.6 K), its pairing symmetry is argued to differ from the \emph{s}$\pm$-wave predicted and observed in many pnictides\cite{Mazin,Kuroki,JPHu,HDing,Hanaguri} due to the existence of gap node(s) reported in thermal conductivity\cite{thermal_SYLi,thermal_Taillefer}, penetration depth\cite{penetration_CHLee}, and nuclear quadrupole resonance\cite{nuclear}. Its superconducting phase may even experience a change of pairing symmetry under pressure as its \emph{T}$\mathrm{_c}$-\emph{P} curve has a minimum at about 1.75 GPa\cite{pressure_Taillefer}. This material exhibits many other unusual properties, including a large normal state Sommerfeld coefficient ($\gamma$$_n$ $\approx$ 100 mJ/mol-K$^2$)\cite{Sommerfeld_Hardy}, a high sensitivity to nonmagnetic impurities (\emph{T}$\mathrm{_c}$ is suppressed to zero with 4\% Co impurity)\cite{impurity_XianHu}, and a huge residual resistivity ratio (RRR) (up to 2000)\cite{impurity_XianHu}, possibly indicating a different superconducting mechanism. %This motivates us to search for novelties in the low-energy electronic structure in both the normal and superconducting states using STS and ARPES, two complementary spectroscopic techniques.
In this work, we report high-resolution STS and ARPES studies of the iron-based superconductor KFe$_{2}$As$_{2}$. We found a vHs just a few meV below \emph{E}$\mathrm{_F}$, which locates in the middle of the principle axes of the first Brillouin zone. The DOS at \emph{E}$\mathrm{_F}$ mainly comes from the vHs while it is non-gapped in the superconducting state. The nodal vHs points provide consistent explanations to many exotic behaviours previously observed in this material and suggests Cooper pairing induced by a strong coupling mechanism.

\section{Experiments}

The single crystals of KFe$_2$As$_2$ were synthesized by using the self-flux method\cite{synthesize}. The starting materials K pieces, Fe powders and As grains (purity 99.9\%, Alfa Aesar) were carefully weighed in an atomic ratio of K:Fe:As=3:2:4 and put into an alumina crucible. Then the crucible was sealed in a tantalum container under 1 atm of argon gas. The proper use of tantalum tube successfully avoids the serious corrosion induced by the potassium vapour. By sealing the container in an evacuated quartz tube, the chemicals were subsequently heated up to 1050$^{\circ}$C and held for 3 hours. Then the furnace was cooled down to 900$^{\circ}$C at a rate of 3$^{\circ}$C/h and at 5$^{\circ}$C/h from 900$^{\circ}$C to 600$^{\circ}$C. Then the sample was cooled down to room temperature by shutting off the power of the furnace. The samples are obtained after removing the leftover KAs flux on the surface of the crystals. Our samples are rather stable in air and look black and shiny.

The scanning tunneling microscopy (STM) and STS were measured in ultrahigh vacuum, low temperature and high magnetic field scanning probe microscope USM-1300 (Unisoku Co., Ltd.). The samples were cleaved at room-temperature with a base pressure about 1$\times$10$^{-10}$ torr, then immediately inserted into the microscope head, which was kept at a low temperature. Etched tungsten tips were used in all STM/STS measurements. The tips were treated by \emph{in-situ} e-beam sputtering and calibrated on a single crystalline Au(100) surface.
A lock-in technique with an ac modulation of 0.3 mV at 987.5 Hz was used to lower down the noise of the differential conductance spectra.

The ARPES measurements were performed at Beamline I05 of Diamond Light Source using a VG-Scienta R4000 electron analyser. The energy and momentum resolutions have been set at 8 meV and 0.2$^{\circ}$, respectively. The samples were cleaved \emph{in situ} and measured at about 7 K in a working vacuum better than 1$\times$10$^{-10}$ torr. The Fermi level \emph{E}$\mathrm{_F}$ of the samples was determined from the crossing of metallic bands in the normal state.

%We fitted our experimental dispersions from ARPES measurements with the two-dimensional tight-binding-like bands. For each band, we integrated   in the \emph{k}$\mathrm{_x}$-\emph{k}$\mathrm{_y}$ 2D Brillouin zone and added them together to get the calculated DOS. We convoluted the DOS with a Gaussian function (FWHM 8 meV) to simulate the experimental resolution. We also calculated a DOS curve without the contribution from the vHs related band, as shown in Fig.~\ref{fig4}(b).

\section{Results}

\begin{figure}
\includegraphics[width=9cm]{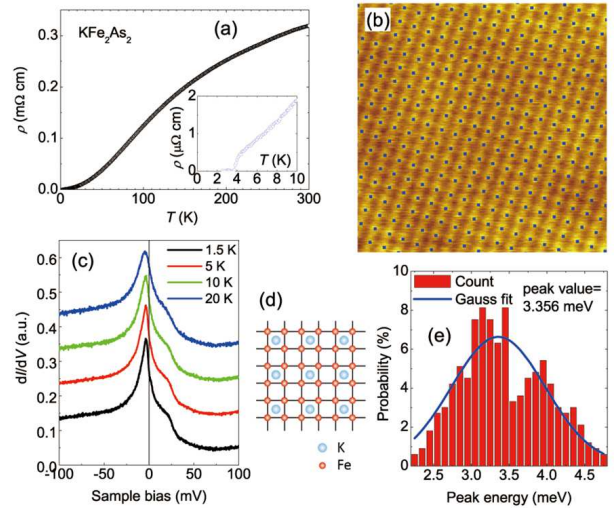}
\caption {(color online) (a) Temperature dependence of resistivity. The inset shows the enlarged view in the low-temperature region of the same data. (b) Topographic STM image (5 nm$\times$5 nm) of the cleaved surface made by K atoms, measured with \emph{V} = 50 mV, and \emph{I}$_t$ = 50 pA. The blue dots here represent the simulated positions of the remaining K atoms. (c) Set of tunneling spectra measured at different temperature with offsets. A sharp peak of \emph{dI/dV} appears at about -3.8 mV. (d) Schematic of the K atoms at the top layer and the Fe atoms beneath the surface layer. (e) Distribution of the vHs peak positions based on a statistics on about 300 STS measured on a 200 nm$\times$200 nm surface at 1.5K. The solid lines shows a Gaussian fit to the distribution.}\label{fig1}
\end{figure}

In Fig.~\ref{fig1}(a), we present the temperature dependence of the electrical resistivity of KFe$_2$As$_2$. The RRR of the KFe$_2$As$_2$ single crystal reaches about 1389, which is very high and suggests good sample quality. Figure~\ref{fig1}(b) shows the atomic image of STM after cleaving. Interestingly, the cleaved surface exhibits a square lattice with a constant of about 5.2 {\AA}. Further analysis indicates that the K atoms reconstruct in an ordered $\sqrt{2}\times\sqrt{2}$ structure of the K atoms in the bulk, maintaining a charge-neutral surface which is very important to measure the proper tunneling curves in the iron-based superconductors\cite{Hoffman}. In Fig.~\ref{fig1}(d), we give a schematic structure of the K atoms at the top layer and the Fe atoms beneath the surface. In some places, we observe some dark void-like defects (not shown here and will be presented separately) which are likely due to Fe vacancies in the layer beneath the surface, as suggested by the STM mapping of the central position of these voids.

Figure~\ref{fig1}(c) displays a set of tunneling spectra measured at 1.5, 5, 10 and 20 K in a defect-free area. One can clearly observe a strong and sharp peak at about -3.8 mV. The spectra exhibit an asymmetric behaviour, with more contribution from the occupied states, which is consistent with situations where the electronic structure is dominated by hole bands\cite{ARPES_Ding,ARPES_Yoshida,ARPES_Ding2}. Through numerous subsequent measurements, we confirm that this type of STS spectrum is a typical one on the cleaved surface, although the peak position is slightly dependent on the measurement position. The statistics on the peak positions of about 300 measured spectra displayed in Fig.~\ref{fig1}(e) show clearly that the peak locates between around 4.5 to 2.5 meV below \emph{E}$\mathrm{_F}$. The appearance of such a sharp peak of \emph{dI/dV} near \emph{E}$\mathrm{_F}$ strongly indicates that there should be one or more very flat bands slightly below \emph{E}$\mathrm{_F}$. From earlier ARPES data\cite{ARPES_Ding}, we can conclude that this cannot appear on anyone of the hole Fermi surface sheets near the Brillouin zone (BZ) centre, the $\Gamma$ (or Z) point. Using our refined ARPES data and simulations, we will demonstrate later that this peak is due to a two-dimensional vHs. From the temperature evolution of the STS, one can see that the peak remains but is slightly broadened above the superconducting transition temperature. Actually, the STS profile at low temperature has a sharp peak at about -3.8 mV with a partially-gapped feature near \emph{E}$\mathrm{_F}$. Because the energy window shown in Fig.~\ref{fig1}(c) is too wide and the data acquisition step is relatively large for this spectrum, one cannot really see a superconducting gap near zero bias energy, but rather a kink. Later we will show that this partially-gapped feature (with $\Delta$ $\approx$ 1.0 mV) disappears above \emph{T}$\mathrm{_c}$ $\approx$ 3.6 K, manifesting that it corresponds to the superconducting gap.

\begin{figure}
\includegraphics[width=9cm]{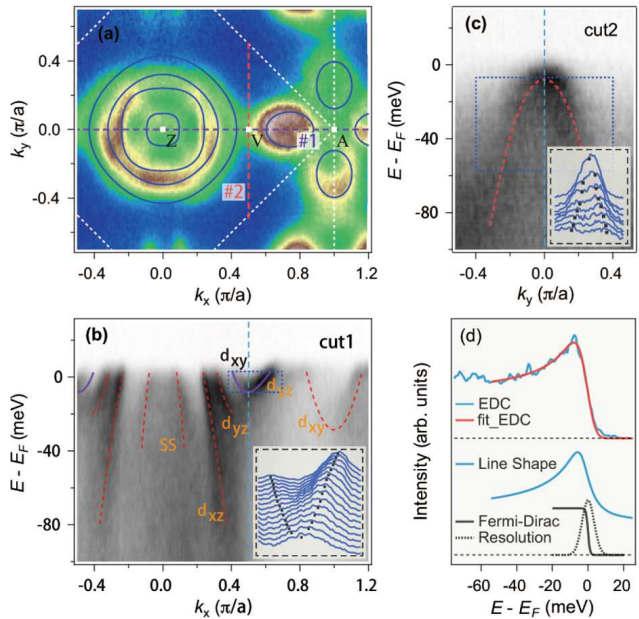}
\caption {(color online) (a) Fermi surface intensity map recorded with 56.5 eV photons (corresponding to \emph{k}$\mathrm{_z}$ = $\pi$). The energy window for integration is 5 meV. The blue curves represent the Fermi surfaces. (b,c) ARPES intensity plot along two mutually perpendicular cuts labelled with \#1 (purple dashed line) and \#2 (red dashed line) in(a), corresponding to the Z-A symmetry line and the (0.5$\pi$/a, \emph{k}$\mathrm{_y}$) line. The electron- and hole-like bands along these two cuts indicate a saddle point at V(0.5$\pi$/a, 0). The insets show MDCs crossing the V point, in the regions emphasized by dashed blue boxes. The orbital characters of the $d$ bands given in (b) are determined from the photon polarization dependence of the ARPES spectrum. Here SS refers to a surface state. (d) Energy distribution curve through the saddle point, indicated as the cyan dashed line in (b) and (c). It is fitted with the asymmetrical line shape proposed in Ref.26, which is multiplied by the Fermi-Dirac function. } \label{fig2}
\end{figure}

Figure~\ref{fig2} shows the ARPES spectra of KFe$_2$As$_2$ measured at 7 K. Regardless of the inner band attributed earlier to a surface state\cite{ARPES_Yoshida}, we observe three hole-like pockets around the zone centre, as illustrated by cut 1 displayed in Fig.~\ref{fig2}(b). In contrast to the Fermi surfaces of moderately K-doped BaFe$_2$As$_2$ and most of the 122-type iron-pnictides, there is no electron-like pocket at the M (or A) point in KFe$_2$As$_2$. Instead, petal shaped hole-like pockets are found, as shown in Fig.~\ref{fig2}(a), which shows the FS intensity map at the \emph{k}$\mathrm{_z}$ = $\pi$ plane. This unique FS topology is consistent with previous studies\cite{ARPES_Ding,ARPES_Yoshida}. A Lifshitz transition occurs in the (Ba,K)Fe$_2$As$_2$ system upon hole-doping\cite{ARPES_Ding2}, which may be responsible for the change of pairing symmetry from \emph{s}-wave to \emph{d}-wave claimed from thermal conductivity data\cite{thermal_Taillefer}.

We now focus our attention to the V($\pi$/2,0) point, located half-way between the $\Gamma$ and M points (or between the Z and A points in the \emph{k}$\mathrm{_z}$ = $\pi$ plane). We show in Figs.~\ref{fig2}(b) and ~\ref{fig2}(c) two mutually perpendicular cuts passing through V, as indicated in Fig.~\ref{fig2}(a). While the band indicated in purple in Fig.~\ref{fig2}(b) is electron-like with a bottom at V, the band dispersion in the perpendicular direction (cut 2), given in Fig.~\ref{fig2}(c), is hole-like with a maximum at V. This observation is also quite clear from the momentum distribution curves (MDCs) in the insets of Fig.~\ref{fig2}(b) and ~\ref{fig2}(c). As a result, there is a saddle point just below \emph{E}$\mathrm{_F}$ at the V point. Since a large portion of this band is of $d_{xy}$ orbital character with almost no dispersion along \emph{k}$\mathrm{_z}$, this saddle point is quite two-dimensional and should produce a logarithmically divergent vHs, thus naturally explaining the strong peak in the STS spectra [Fig.~\ref{fig1}(c)]. Indeed, as shown in Fig.~\ref{fig2}(d), a peak in the EDC (energy distribution curve) at the V point located a few meV below \emph{E}$\mathrm{_F}$ is observed by ARPES, almost at the same energy as the DOS peak observed by STS. The EDC curve is also fitted with the asymmetrical lineshape proposed in Ref. 26, which is multiplied by the Fermi-Dirac function. We caution that the vHs observed at the V point is fundamentally different from the tops and bottoms of bands that have been proposed previously to play a role in iron-based superconductivity\cite{vHs_iron}, the latter can also be called as a vHs in a general terminology.

\begin{figure}
\includegraphics[width=9cm]{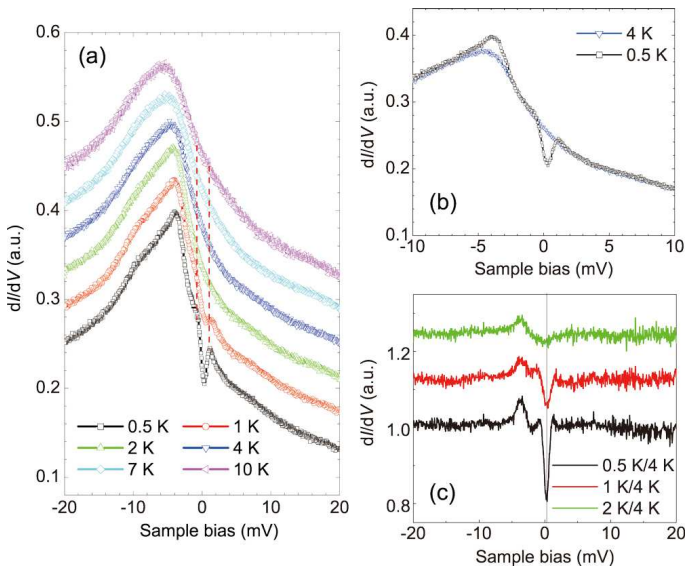}
\caption {(color online) (a) STS spectra measured from 0.5 K to 10 K with offsets. One can see that the vHs peak and the superconducting gap are both observed at low temperatures. Above \emph{T}$\mathrm{_c}$, the superconducting gap disappears but the vHs peak remains and gets slightly broadened. (b) STS spectra measured at $T = 0.5$ K and 4 K. The background of the two spectra look very similar. (c) STS spectrum measured at $T = 0.5$ K, 1.0 and 2.0 K divided by that measured at $T =$ 4.0 K. A superconducting gap with $\Delta\approx1$ meV can be clearly seen, but the depleted DOS due to the superconducting gap is very small, leading to a large ungapped contribution arising from the vHs.} \label{fig3}
\end{figure}

In Fig.~\ref{fig3}, we illustrate how the STS profile with a vHs peak and the superconducting gap evolve with temperature. The STS were taken at a point far away from the defects. As demonstrated in Fig.~\ref{fig3}(a), the vHs peak is very sharp in the STS spectra measured at very low temperatures, and the superconducting gap is also clearly visible. With increasing temperature, the superconducting gap disappears above \emph{T}$\mathrm{_c}$, while the vHs peak remains but is slightly broadened. Due to this broadening effect, the peak position shifts from about -3.9 mV to about -5.4 mV when temperature is increased from 0.5 K to 10 K. In Fig.~\ref{fig3}(b), we overlap two STS spectra measured at \emph{T} = 0.5 K and 4.0 K, respectively, in a narrow energy window. We clearly observe a superconducting gap on the curve measured at 0.5 K. Worthy of noting is that the two spectra have quite similar background, with a vHs peak in both the superconducting and the normal states. From the raw data in Figs.~\ref{fig3}(a) and ~\ref{fig3}(b), we find that the loss of DOS due to superconductivity at low temperature is quite limited. To evaluate the gapped DOS, we divide the STS spectrum measured at \emph{T} = 0.5, 1 and 2 K by that recorded at \emph{T} = 4.0 K. This leads to a clear gapped feature, but with a quite high ungapped DOS near \emph{E}$\mathrm{_F}$, as shown in Fig.~\ref{fig3}(c). The anomalous peaks at around -4 meV in Fig.~\ref{fig3}(c) is induced by the variation of the vHs with temperature, which is not the focus here. From the normalized data of STS at 0.5 K, we determined a gap value of about 1 meV. The laser ARPES\cite{ARPES_Shin} data show several anisotropic gaps with gap maximum of 1 meV on the inner hole Fermi pocket around Gamma point at 2 K. The exact gap value from the STS spectra is influenced by the gap structure and magnitude, and usually smaller than the value determined by the distance between two coherence peaks, so the gap value from two kinds of measurements are comparable.
The suppression of the DOS at zero energy due to superconductivity is about 20\%, thus leaving the major part of the DOS at \emph{E}$\mathrm{_F}$ ungapped. Similar spectra have been observed and repeated as always at other locations distant away from the defects, which suggests that this ungapped DOS is not due to impurity scattering. An additional argument for us to rule out the impurity scattering as the cause for the ungapped DOS is that the sample is very clean with a very large RRR. In contrast to our expectation, the superconducting gap feature becomes already weak when it is measured at 2 K. In conventional superconductors with \emph{s}-wave gaps, the gap features are still very clear when measured at about \emph{t}=\emph{T}$\mathrm{_c}$/2. This can be self-consistently explained by the fact that in the temperature range investigated here the system contains a dilute superfluid density and a large number of unpaired electrons.

\begin{figure}
\includegraphics[width=9cm]{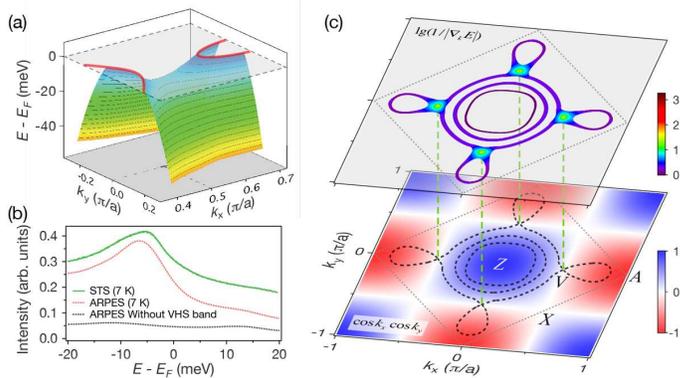}
\caption {(color online) (a) Three-dimensional representation of the band dispersion around the saddle point fitted from the ARPES data. (b) Comparison between the calculated DOS obtained from ARPES band dispersions and the DOS measured from STS experiments. The STS spectrum (green line) has been shifted up by 0.05 for clarity. (c) Upper panel: Intensity plot of log($1/|\nabla_kE|$) derived from the ARPES experimental band dispersion, which corresponds to the $k$-dependent contributions to the DOS. We integrate this quantity within $\pm$5 meV from the saddle point. A bright spot corresponding to high DOS is observed at each V location. Lower panel: Energy contour at the binding energy of the saddle points (dashed curves) on the image of the extended $s$-wave gap function . The saddle points are exactly located at the nodal lines marked by the white band.} \label{fig4}
\end{figure}

In Fig.~\ref{fig4}(a), we show a three-dimensional representation of the energy band near V with the saddle point fitted from the ARPES data with the two-dimensional tight-binding-like bands. For each band, we integrated in the \emph{k}$\mathrm{_x}$-\emph{k}$\mathrm{_y}$ 2D Brillouin zone and added them together to get the calculated DOS. We convoluted the DOS with a Gaussian function (FWHM 8 meV) to simulate the experimental resolution. We also calculated a DOS curve without the contribution from the vHs related band. Figure~\ref{fig4}(b) compares the calculated DOS obtained from ARPES band dispersions and the measured DOS from STS measurements, and a remarkable similarity between the two DOS curves is observed. Interestingly, we note that approximately 80\% of the states at \emph{E}$\mathrm{_F}$ in our calculated DOS originate from the four saddle points within one BZ. Since 80\% of the states at \emph{E}$\mathrm{_F}$ measured by STS are ungapped, it is reasonable to assume that the vHs and its vicinity in the momentum space are mostly ungapped. To visualize the distribution of the DOS in the BZ, we plot $1/|\nabla_kE|$ using a logarithmic scale in the upper panel of Fig.~\ref{fig4}(c). We integrate this quantity within $\pm$5 meV from the saddle point for this plot. A bright spot is observed at each V location, with the intensity several orders of magnitude higher than anywhere else.

\section{Discussion}

The existence of a strong vHs in KFe$_2$As$_2$ is quite striking and may offer a consistent explanation to many surprising properties previously observed in this material. First, the existence of the four vHs points naturally explains the large enhancement of the Sommerfeld coefficient in the normal state\cite{Sommerfeld_Hardy}.
One can see that there is a turning point of electronic specific heat \emph{C}$\mathrm{_e}$/\emph{T} just at 0.5 K, below which a sharper drop appears. This is consistent with our observation that the large DOS at \emph{E}$\mathrm{_F}$ is contributed by the vHs but it is ungapped at and above 0.5 K. Therefore, one can naturally conclude that the vHs plays only a passive role and is not the driving force of the pairing. Our results are also qualitatively consistent with de Haas van Alphen measurements suggesting that the highest DOS is coming from the most outer Fermi surface around the $\Gamma$ point, namely the $\beta$ Fermi surface as defined in our paper\cite{de Haas}, while the largest superconducting gap appears at the inner hole pocket\cite{ARPES_Shin}. It is also important to note that heavy mass has been reported\cite{TlNi2Se2_1,TlNi2Se2_2} in TlNi$_2$Se$_2$, which is isostructural to KFe$_2$As$_2$ and for which ARPES also revealed a vHs near \emph{E}$\mathrm{_F}$. In addition, the observation of the gapless behaviour in the DOS near the vHs explains the origin of the nodal gap observed in this material\cite{thermal_SYLi,thermal_Taillefer,penetration_CHLee,nuclear}. Finally, the unusual pressure dependence of superconducting transition temperature\cite{pressure_Taillefer} can possibly be understood as a Lifshitz transition associated with the vHs. We note that such a Lifshitz transition will not change the Fermi surface volume by much since both Fermi surface sheets connected by the vHs are hole-like. Since there is no superconducting gap contributed from the vHs point, the observed \emph{T}$\mathrm{_c}$ minimum under pressure can be consistently interpreted as correlated with a Lifshitz transition.

Our observation of the low-energy vHs in KFe$_2$As$_2$ may have even more profound implications to the interplay between the vHs and superconductivity because such kind of vHs may enhance the instability towards superconductivity with unconventional pairing symmetries. Our data provide a strong constraint on the candidates of possible emergent pairing symmetries due to the vHs. Despite the high DOS near \emph{E}$\mathrm{_F}$ due to the vHs, this DOS is however not gapped at a temperature of 0.5 K, which is already quite low compared with \emph{T}$\mathrm{_c}$. This is certainly counter-intuitive to the pairing mechanism based on the weak coupling picture because according to that scenario the superconducting gap can be expressed as $\Delta_{sc}\sim$ exp$(-1/N_FV_{int})$, with DOS $N_F$ at the Fermi energy, \emph{$V_{int}$} the pairing interaction mediated by exchanging bosons. Therefore it is difficult to accept a simple mechanism based on the weak coupling picture for the pairing. In addition, two candidates for the gap symmetry have been popularly proposed in KFe$_2$As$_2$. One is the $d_{x^2-y^2}$ wave argued earlier\cite{pressure_Taillefer} and the other is \emph{p}$_x$+\emph{ip}$_y$, which is possible because of the existence of the four type-II vHs points in one BZ\cite{BiS2_1,BiS2_2}. Unfortunately, both pairing symmetries are inconsistent with our data because the superconducting gaps at the vHs point should be enhanced in both cases. However, our results are consistent with the scenario in which the $s\pm$ pairing remains robust and unaffected by the vHs in such a heavily hole-doped material if we consider that the corresponding leading gap function is cos\emph{k}$\mathrm{_x}$cos\emph{k}$\mathrm{_y}$\cite{JPHu,J.P.Hu}, which has nodal lines going through the four vHs points, as illustrated in Fig.~\ref{fig4}(c). Although the vHs occurs at about -3.8 meV from the Fermi energy, but it is quite close to the latter, the finite broadness of the band dispersion can still give a finite DOS at $E_F$, which contributes the ungapped DOS near the momentum $V$ point when the gap function is cos\emph{k}$\mathrm{_x}$cos\emph{k}$\mathrm{_y}$. This scenario is further supported by our earlier ARPES superconducting gap measurements in Ba$_{0.1}$K$_{0.9}$Fe$_2$As$_2$ (\emph{T}$\mathrm{_c}$ = 9 K) \cite{ARPES_Ding2}. If we assume that the vHs induces an additional unconventional pairing that coexists with the s$\pm$ pairing, the likely candidate is $d_{xy}$, which also has nodal lines going through the four vHs points.

\section{Conclusions}
In summary, we perform STS and ARPES studies on the iron-based superconductor KFe$_{2}$As$_{2}$. Our observation of a pronounced vHs with large ungapped DOS in the vicinity of \emph{E}$\mathrm{_F}$ in KFe$_2$As$_2$ points to the importance of the strong coupling mechanism. This work not only posts new challenges for the theoretical understanding of the interplay between vHs and superconductivity, but also adds new ingredients and new physics to the already rich properties of the iron-based superconductors.

\begin{acknowledgments}
We thank Dunghai Lee, Wei Ku, Zi-Yang Meng and Sergey Borisenko for valuable discussions. This work was supported by the Ministry of Science and Technology of China (973 projects: 2011CBA00102, 2010CB923000, 2012CB821403), NSF of China (11004232, 11034011, 11234014 and 11274362), the Strategic Priority Research Program of the CAS (XDB07000000) and PAPD.

\end{acknowledgments}

$^{\dag}$dingh@iphy.ac.cn, $^*$hhwen@nju.edu.cn

\end{document}